\documentclass[10pt,journal,compsoc]{IEEEtran}
\usepackage{amsmath}
\usepackage{amssymb}
\usepackage{mathtools}
\usepackage{url}
\usepackage{graphicx}
\usepackage{listings}
\usepackage{subfig}
\usepackage{color}
\usepackage{booktabs} 
\usepackage{fancyvrb}
\usepackage{algorithmic}
\usepackage{multirow}
\usepackage[ruled]{algorithm2e} 

\SetAlFnt{\small}
\SetAlCapFnt{\small}
\SetAlCapNameFnt{\small}
\SetAlCapHSkip{0pt}
\IncMargin{-\parindent}

\providecommand{\changed}{}

\ifCLASSOPTIONcompsoc
  \usepackage[nocompress]{cite}
\else
  \usepackage{cite}
\fi
\ifCLASSINFOpdf
\else
\fi
\begin{document}

%
\title{SLEEF: A Portable Vectorized Library of\\C Standard Mathematical Functions}
%
%
%
%

\author{Naoki~Shibata,~\IEEEmembership{Member,~IEEE,}
        and~Francesco~Petrogalli
\IEEEcompsocitemizethanks{\IEEEcompsocthanksitem Naoki Shibata is with
  Graduate School of Information Science, Nara
  Institute of Science and Technology, Nara, Japan.}
\IEEEcompsocitemizethanks{\IEEEcompsocthanksitem Francesco Petrogalli is with
  ARM, 110 Fulbourn Road, Cambridge, CB1 9NJ, United Kingdom.}
\thanks{Manuscript received 25 Apr. 2019; revised 11 Dec. 2019; accepted 13 Dec. 2019. Date of publication 18 Dec. 2019.}
\thanks{(Correcponding author: Naoki Shibata.)}
\thanks{Recommended for acceptance by D. S. Nikolopoulos.}
\thanks{Digital Object Identifier no. 10.1109/TPDS.2019.2960333}}

%
%

\markboth{IEEE Transactions on PARALLEL AND DISTRIBUTED SYSTEMS,~Vol.~31, No.~6, JUNE~2020}%
{Shell \MakeLowercase{\textit{et al.}}: SLEEF: A Portable Vectorized Library of C Standard Mathematical Functions}
%



\lstset{
  language=C,
  escapeinside={/*}{*/},
  showstringspaces=false,
  columns=flexible,
  basicstyle={\footnotesize\ttfamily},
  numbers=left,
  numbersep=5pt,
  xleftmargin=5pt,
  xrightmargin=5pt,
  breaklines=true,
  breakatwhitespace=true,
  tabsize=4
}

\IEEEtitleabstractindextext{%
\begin{abstract}
  In this paper, we present techniques used to implement our portable
  vectorized library of C standard mathematical functions written
  entirely in C language. In order to make the library portable while
  maintaining good performance, intrinsic functions of vector
  extensions are abstracted by inline functions or preprocessor
  macros. We implemented the functions so that they can use
  sub-features of vector extensions such as fused multiply-add, mask
  registers and extraction of mantissa. In order to make computation
  with SIMD instructions efficient, the library only uses a small
  number of conditional branches, and all the computation paths are
  vectorized. We devised a variation of the Payne-Hanek argument
  reduction for trigonometric functions and a floating point
  remainder, both of which are suitable for vector computation. We compare the
  performance of our library to Intel SVML.
\end{abstract}

\begin{IEEEkeywords}
Parallel and vector implementations, SIMD processors, elementary
functions, floating-point arithmetic
\end{IEEEkeywords}}

\maketitle

\IEEEdisplaynontitleabstractindextext

%
\IEEEpeerreviewmaketitle

\IEEEraisesectionheading{\section{Introduction}\label{sec:introduction}}

\IEEEPARstart{T}{he} instruction set architecture of most modern
processors provides \emph{Single Instruction Multiple Data (SIMD)}
instructions that process multiple instances of data concurrently{\cite{875989}}. The
programming model that utilizes these instructions is a key technique for
many computing systems to reach their peak performance. Most
software SIMD optimizations are introduced manually by
programmers. However, this approach introduces a portability problem
because the code needs to be re-written when targeting a new vector
extension. In order to improve portability of codes with SIMD
optimizations, recent compilers have introduced auto-vectorizing
capability{\cite{naishlos-2004}}. To fully exploit the SIMD capabilities of a
system, the transformation for auto-vectorization of a compiler must
be able to invoke a version of functions that operates on
concurrent iterations, or on a \emph{vector function}. This applies particularly to C
mathematical functions defined in \texttt{math.h} that are frequently called in
hot-loops.

In this paper, we describe our implementation of a vectorized library of
C standard math functions, called SLEEF library. SLEEF stands for
\emph{SIMD Library for Evaluating Elementary Functions}, and
implements a vectorized version of all C99 real floating-point math
functions. Our library provides 1-ULP accuracy version and 3.5-ULP
accuracy version for most of the functions. We confirmed that our library satisfies such accuracy requirements
on an empirical basis. Our library achieves both good performance and portability
by abstracting intrinsic functions. This abstraction enables
sub-features of vector extensions such as mask registers to be utilized
while the source code of our library is shared among different vector
extensions. We also implemented a
version of functions that returns bit-wise consistent results across
all platforms. Our library is designed to be used in conjunction with
vectorizing compilers.
In order to help development of vectorizing
compilers, we collaborated with compiler developers in designing a
Vector Function Application Binary Interface (ABI). The main difficulty in vectorizing math functions is that
conditional branches are expensive.
We implemented many of the
functions in our library without conditional branches. We devised
reduction methods and adjusted domains of polynomials so that a single
polynomial covers the entire input domain.
For an increased vector size, a value requiring a slow
  path is more likely to be contained in a vector. Therefore, we
  vectorized all the code paths in order to speed up the computation
  in such cases.
We devised a variation of the Payne-Hanek range
reduction and a remainder calculation method that are both suitable
for vectorized implementation.

We compare the implementation of several selected functions in our
library to those in other open-source libraries. We also compare
the reciprocal throughput of functions in our library, Intel SVML~\cite{SVML}, FDLIBM~\cite{FDLIBM}, and
Vector-libm~\cite{7869070}. We show that the performance of our library is comparable
to that of Intel SVML.

The rest of this paper is organized as follows.
Section~\ref{sec:related} introduces related work. Section~\ref{sec:veal}
discusses how portability is improved by abstracting
vector extensions. Section~\ref{sec:background} explains the
development of a Vector ABI and a vectorized mathematical
library. Section~\ref{sec:overview} shows an overview of the
implementation of SLEEF, while comparing our library with FDLIBM and
Vector-libm. Section~\ref{sec:testing} explains how our library is
tested.  Section~\ref{sec:performance} compares our work with prior
art. In Section~\ref{sec:conclusion}, the conclusions are presented.

\section{Related Work}

\label{sec:related}

\subsection{C Standard Math Library}

The C standard library (libc) includes the standard mathematical
library (libm)~\cite{ISO:2011:IIIb}. There have been many
implementations of libm. Among them, FDLIBM~\cite{FDLIBM} and the libm
included in the GNU C Library~\cite{glibc} are the most widely used
libraries. FDLIBM is a freely distributable libm developed by Sun
Microsystems, Inc., and there are many derivations of this library.
Gal et al. described the algorithms used in the elementary
mathematical library of the IBM Israel Scientific
Center~\cite{Gal:1991:AEM:103147.103151}. Their algorithms are based
on the accurate tables method developed by Gal. It achieves high
performance and produces very accurate results. Crlibm is a project to
build a correctly rounded mathematical library~\cite{CRLIBM}.

There are several existing vectorized implementations of libm. Intel
Short Vector Math Library (SVML) is a highly regarded commercial
library~\cite{SVML}. This library provides highly optimized
subroutines for evaluating elementary functions which can use several
kinds of vector extensions available in Intel's processors.  However,
this library is proprietary and only optimized for Intel's
processors. There are also a few commercial and open-source
implementations of vectorized libm. AMD is providing a vectorized libm
called AMD Core Math Library (ACML)~\cite{ACML}.

Some of the code from SVML is published under a free software license,
and it is now published as Libmvec~\cite{libmvec}, which is a part of
Glibc. This library provides functions with 4-ULP error bound. It is
coded in assembly language, and therefore it does not have good
portability. C.~K.~Anand et~al. reported their C implementation of 32
single precision libm functions tuned for the Cell BE SPU compute
engine~\cite{4731241}. They used an environment called Coconut that
enables rapid prototyping of patterns, rapid unit testing of assembly
language fragments and patterns to develop their library.
M.~Dukhan published an open-source and
portable SIMD vector libm library named Yeppp!~\cite{yeppp,
  Dukhan2014}. Most of vectorized implementations of libm utilizes
assembly coding or intrinsic functions to specify which vector
instruction is used for each operator. On the other hand, there are
also other implementations of vector versions of libm which are written in a scalar
fashion but rely on a vectorizing compiler to generate vector instructions and 
generate a vectorized binary code. Christoph Lauter published an
open-source Vector-libm library implemented with plain
C~\cite{7869070}. VDT Mathematical
Library~\cite{1742-6596-513-5-052027}, is a math library written for
the compiler's auto-vectorization feature.

\subsection{Translation of SIMD Instructions}

Manilov et~al. propose a C source code translator for substituting
calls to platform-specific intrinsic functions in a source code with
those available on the target
machine~\cite{Manilov:2016:FRR:3025020.2990194}. This technique
utilizes graph-based pattern matching to substitute intrinsics. It can
translate SIMD intrinsics between extensions with different vector
lengths. This rewriting is carried out through loop-unrolling.

N.~Gross proposes specialized C++ templates for making the source code
easily portable among different vector extensions without sacrificing
performance~\cite{7568423}. With
these templates, some part of the source code can be written in a way
that resembles scalar code. In order to vectorize algorithms that have a
lot of control flow, this scheme requires the bucketing technique is
applied, to compute all the paths and choose the relevant results at the end.

Clark et~al. proposes a method for combining static analysis at
compile time and binary translation with a JIT compiler in order to
translate SIMD instructions into those that are available on the target
machine~\cite{4147662}. In this method, SIMD instructions in the code
are first converted into an equivalent scalar representation. Then, a
dynamic translation phase turns the scalar representation back into
architecture-specific SIMD equivalents.

Lei{\ss}a et~al. propose a C-like language for portable and efficient
SIMD programming~\cite{LeiBa:2012:ECL:2370036.2145825}. With their
extension, writing vectorized code is almost as easy as writing
traditional scalar code. There is no strict separation in host code
and kernels, and scalar and vector programming can be mixed. Switching
between them is triggered by the type system. The authors present a
formal semantics of their extension and prove the soundness of the
type system.

Most of the existing methods are aiming at translating SIMD intrinsics
or instructions to those provided by a different vector extension in
order to port a code. Intrinsics that are unique in a specific
extension are not easy to handle, and translation works only if
the source and the target architectures have equivalent SIMD
instructions. Automatic vectorizers in compilers have a similar
weakness. Whenever possible, we have specialized the implementation of
the math functions to exploit the SIMD instructions that are specific
to a target vector extension. We also want to make special
handling of FMA, rounding and a few other kinds of instructions, because
these are critical for both execution speed and accuracy. We want to
implement a library that is statically optimized and usable with
Link Time Optimization (LTO). The users of our library do not appreciate usage of a JIT
compiler. In order to minimize dependency on external libraries, we
want to write our library in C. In order to fulfill these requirements,
we take a cross-layer approach. We have been developing our abstraction
layer of intrinsics, the library implementation, and the algorithms in
order to make our library run fast with any vector extensions.

\section{Abstraction of Vector Extensions}

\label{sec:veal}

Modern processors supporting SIMD instructions have SIMD
registers that can contain multiple data~\cite{875989}. For example, a
128-bit wide SIMD register may contain four 32-bit single-precision FP
numbers. A SIMD add instruction might take two of these registers as
operands, add the four pairs of numbers, and overwrite one of these
registers with the resulting four numbers. We call an array of FP numbers
contained in a SIMD register a vector.

SIMD registers and instruction can be exposed in a C program with 
intrinsic functions and types~\cite{armmanual}. An intrinsic
function is a kind of inline function that exposes the architectural
features of an instruction set at C level. By calling an intrinsic
function, a programmer can make a compiler generate a specific
instruction without hand-coded assembly. Nevertheless, the compiler
can reorder instructions and allocate registers, and therefore
optimize the code. When intrinsic functions corresponding to
SIMD instructions are defined inside a compiler, C data types for
representing vectors are also defined.

In SLEEF, we use intrinsic functions to specify which assembly
instruction to use for each operator. We abstract intrinsic functions for each vector extension
by a set of inline functions or preprocessor macros. We implement the functions
exported from the library to call abstract
intrinsic functions instead of directly calling intrinsic
functions. In this way, it is easy to swap the vector extension to
use. We call our set of inline functions for abstracting
architecture-specific intrinsics \emph{Vector Extension Abstraction
  Layer (VEAL)}. 


In some of the existing vector math libraries, functions are
implemented with hand-coded assembly\cite{libmvec}. This approach improves the
absolute performance because it is possible to provide the optimal
implementation for each microarchitecture. However, processors with a
new microarchitecture are released every few years, and the library
needs revision accordingly in order to maintain the optimal
performance.

In other vector math libraries, the source code is written in
a scalar fashion that is easy for compilers to
auto-vectorize\cite{7869070, 1742-6596-513-5-052027}. Although such libraries have good portability, it is
not easy for compilers to generate a well-optimized code. In order for
each transformation rule in an optimizer to kick in, the source code
must satisfy many conditions to guarantee that the optimized code runs
correctly and faster. In order to control the level of optimization,
a programmer must specify special attributes and compiler
options.

\subsection{Using Sub-features of the Vector Extensions}

\begin{figure}[t]
\begin{tabular}{c}
\begin{lstlisting}
// Type definition
typedef svbool_t vopmask;
typedef svfloat64_t vdouble;

// Abstraction of intrinsic functions
#define ptrue svptrue_b8()
vdouble vcast_vd_d(double d) { return svdup_n_f64(d); }
vdouble vsub_vd_vd_vd(vdouble x, vdouble y) {
  return svsub_f64_x(ptrue, x, y); }
vopmask veq_vo_vd_vd(vdouble x, vdouble y) {
  return svcmpeq_f64(ptrue, x, y); }
vopmask vlt_vo_vd_vd(vdouble x, vdouble y) {
  return svcmplt_f64(ptrue, x, y); }
vopmask vor_vo_vo_vo(vopmask x, vopmask y) {
  return svorr_b_z(ptrue, x, y); }
vdouble vsel_vd_vo_vd_vd(vopmask mask, vdouble x, vdouble y) {
  return svsel_f64(o, x, y); }
\end{lstlisting}
\end{tabular}
\caption{A part of definitions in VEAL for SVE}
\label{fig:veal}
\end{figure}

\begin{figure}[t]
\begin{tabular}{c}
\begin{lstlisting}
vdouble xfdim(vdouble x, vdouble y) {
  vdouble ret = vsub_vd_vd_vd(x, y);
  ret = vsel_vd_vo_vd_vd(vor_vo_vo_vo(vlt_vo_vd_vd(ret, vcast_vd_d(0)), veq_vo_vd_vd(x, y)),
			 vcast_vd_d(0),
			 ret);
  return ret;
}
\end{lstlisting}
\end{tabular}
\caption{Implementation of vectorized {\tt fdim} (positive difference) function with VEAL}
\label{fig:xfdim}
\end{figure}

\label{sec:opmask}

There are differences in the features provided by different vector
extensions, and we must change the function
implementation according to the available features. Thanks to the level of abstraction provided by the VEALs, we
implemented the functions so that all the different
versions of functions can be built from the same source files with different macros enabled.
For example, the availability of FMA instructions is
important when implementing double-double (DD)
operators~\cite{Muller:1997:EFA:261217}. We
implemented DD operators both with and without FMA by 
manually specifying if the compiler can convert each combination of multiplication and
addition instructions to an FMA instruction, utilizing VEALs.

Generally, bit masks are used in a vectorized code in order to conditionally choose elements from two vectors.
In some vector extensions, a
vector register with a width that matches a vector register for
storing FP values, is used to store a bit mask. Some vector extensions provide narrower vector
registers that are dedicated to this purpose, which is
SLEEF makes use of these opmask registers by providing a
dedicated data type in VEALs. If a vector extension does not support an opmask,
the usual bit mask is used instead of an opmask. It is also better to
have an opmask as an argument of a whole math function and make that
function only compute the elements specified by the opmask. By
utilizing a VEAL, it is also easy to implement such a functionality.

\subsection{Details of VEALs}

Fig.~\ref{fig:veal} shows some definitions in the VEAL for
SVE~\cite{sveacle}. We abstract vector data types and intrinsic
functions with typedef statements and inline functions, respectively.

The {\tt vdouble} data type is for storing vectors of double precision FP
numbers. {\tt vopmask} is the data type for the opmask described in
\ref{sec:opmask}.

The function {\tt vcast\_vd\_d} is a function that returns a vector in
which the given scalar value is copied to all elements in the vector.
{\tt vsub\_vd\_vd\_vd} is a function for vector subtraction between
two vdouble data. {\tt veq\_vo\_vd\_vd} compares elements of two
vectors of vdouble type. The results of the comparison can be used, 
for example, by {\tt vsel\_vd\_vo\_vd\_vd} to choose a value for each
element between two vector registers. Fig.~\ref{fig:xfdim} shows an
implementation of a vectorized positive difference function using a
VEAL. This function is a vectorized implementation of the {\tt fdim}
function in the C standard math library.

\subsection{Making Results Bit-wise Consistent across All Platforms}

The method of implementing math functions described so far, can
deliver computation results that slightly differ depending on
architectures and other conditions, although they all satisfy the
accuracy requirements, and other specifications. However, in some applications, bit-wise
consistent results are required.

To this extent, the SLEEF project has been working closely with Unity
Technologies,\footnote{\url{https://unity3d.com/}.} which specializes
in developing frameworks for video gaming, and we discovered
that they have unique requirements for the functionalities of math
libraries. Networked video games run
on many gaming consoles with different architectures and they share
the same virtual environment. Consistent results of simulation at each
terminal and server are required to ensure fairness among all
players. For this purpose, fast computation is more important than
accurate computation, while the results of computation have to
perfectly agree between many computing nodes, which are not guaranteed
to rely on the same architecture. Usually, fixed-point
arithmetic is used for a purpose like this, however there is a demand
for modifying existing codes with FP computation to support
networking.

There are also other kinds of simulation in which bit-wise
identical reproducibility is important. In \cite{climate}, the authors
show that modeled mean climate states, variability and trends at
different scales may be significantly changed or even lead to opposing
results due to the round-off errors in climate system
simulations. Since reproducibility is a fundamental principle of
scientific research, they propose to promote bit-wise identical
reproducibility as a worldwide standard.


One way to obtain bit-wise consistent values from math functions is to
compute correctly rounded values. However, for applications like
networked video games, this might be too expensive. SLEEF provides
vectorized math functions that return bit-wise consistent results
across all platforms and other settings, and this is also achieved by
utilizing VEALs. The basic idea is to always apply the same sequence
of operations to the arguments. The IEEE 754 standard guarantees 
that the basic arithmetic operators give correctly rounded results~\cite{Muller:2009:HFA:1823389},
and therefore the results from these operators are bit-wise consistent. Because
most of the functions except trigonometric functions do not have a
conditional branch in our library, producing bit-wise consistent results is fairly
straightforward with VEALs. Availability of FMA instructions is another key for
making results bit-wise consistent. Since FMA instructions are critical for
performance, we cannot just give up using FMA instructions. In SLEEF,
the bit-wise consistent versions of functions have two versions both with
and without FMA instructions. We provide a non-FMA version of the functions to guarantee bit-wise
consistency among extensions such as Intel SSE2 that do not have FMA instructions.
Another issue is that the compiler might introduce inconsistency by FP
contraction, which is the result of combining a pair of multiplication and addition operations
into an FMA. By disabling FP contraction, the compiler strictly preserves the order and the type of FP
operations during optimization. It is also important to make the returned values from
scalar functions bit-wise consistent with the vector functions. In
order to achieve this, we also made a VEAL that only uses scalar
operators and data types. The bit-wise consistent and non-consistent
versions of vector and scalar functions are all built from the same
source files, with different VEALs and macros enabled.
As described in Section~\ref{sec:overview}, trigonometric functions in
SLEEF chooses a reduction method according to the maximum argument of
all elements in the argument vector. In order to make the returned
value bit-wise consistent, the bit-wise consistent version of the
functions first applies the reduction method for small arguments to
the elements covered by this method. Then it applies the second method
only to the elements with larger arguments which the first
  method does not cover.

\section{The Development of a Vector Function ABI and SLEEF}

\label{sec:background}

Recent compilers are developing new optimization techniques to
automatically vectorize a code written in standard programming
languages that do not support parallelization~\cite{krizikalla,5764683}.
Although the first SIMD and vector
computing systems~\cite{Russell:1978:CCS:359327.359336} appeared a few
decades ago, compilers with auto-vectorization capability have not
been widely used until recently, because of several difficulties in
implementing such functionality for modern SIMD architectures. Such
difficulties include verifying whether the compiler can vectorize a loop or
not, by determining data access patterns of the operations in the
loop~\cite{naishlos-2004,Nuzman:2006:AID:1133981.1133997}.
For languages like C and C++, it is also difficult to determine the
data dependencies through the iteration space of the loop, because it
is hard to determine aliasing conditions of the arrays processed in
the loop.

\subsection{Vector Function Application Binary Interface}

Vectorizing compilers convert calls to scalar versions of math
functions such as sine and exponential to the SIMD version of the
math functions. The most recent versions of Intel
Compiler~\cite{Tian2002IntelOC}, GNU Compiler~\cite{gccweb}, and Arm Compiler for HPC~\cite{armclangweb}, which is based on
Clang/LLVM~\cite{llvmweb, LLVM:CGO04}, are capable of this transformation, and rely on the
availability of vector math libraries such as SVML~\cite{SVML},
Libmvec~\cite{libmvec} and SLEEF respectively to provide an
implementation of the vector function calls that they generate. In
order to develop this kind of transformations, a target-dependent
\emph{Application Binary Interface (ABI)} for calling vectorized
functions had to be designed.

The Vector Function ABI for AArch64 architecture~\cite{vabi-aarch64}
was designed in close relationship with the development of SLEEF. This
type of ABI must standardize the mapping between scalar
functions and vector functions. The existence of a standard enables
interoperability across different compilers, linkers and libraries,
thanks to the use of standard names defined by the specification.

The ABI includes a name mangling function, a map that converts the
scalar signature to the vector one, and the calling conventions that
the vector functions must obey. In particular, the name mangling
function that takes the name of the scalar function to the vector
function must encode all the information that is necessary to reverse
the transformation back to the original scalar function. A linker can use this reverse
mapping to enable more optimizations (Link Time
Optimizations) that operate on object files, and does not have access
to the scalar and vector function prototypes. There is a demand by
users for using a different vector math library according to the
usage. Reverse mapping is also handy for this purpose. A vector math
library implements a function for each combination of a vector
extension, a vector length and a math function to evaluate. As a
result, the library exports a large number of functions. Some
vector math libraries can only implement part of all the
combinations. By using the reverse mapping mechanism, the compiler can check the availability of
the functions by scanning the symbols exported by a
library.

The Vector Function ABI is also used with OpenMP~\cite{OpenMP4}. From
version 4.0 onwards, OpenMP provides the directive \emph{declare
  simd}. A user can decorate a function with this directive to inform
the compiler that the function can be safely invoked concurrently on
multiple instances of its
arguments~\cite{10.1007/978-3-319-65578-9_5}. This means that the compiler can vectorize the
function safely. This is particularly useful when
the function is provided via a separate module, or an external
library, for example in situations where the compiler is not able to
examine the behavior of the function in the call site. The
scalar-to-vector function mapping rules stipulated in the Vector
Function ABI are based on the classification of vector functions
associated with the \texttt{declare simd} directive of OpenMP.
Currently, work for implementing these OpenMP directives on LLVM is
ongoing.

The Vector Function ABI specifications are provided for the Intel x86
and the Armv8 (AArch64) families of vector
extensions~\cite{vabi-x86,vabi-aarch64}. The compiler generates SIMD function
calls according to the compiler flags. For example, when
targeting AArch64 SVE auto-vectorization, the compiler will transform
a call to the standard \texttt{sin} function to a call to the symbol
\texttt{\_ZGVsMxv\_sin}. When targeting Intel AVX-512~\cite{7453080}
auto-vectorization, the compiler would generate a call to the symbol
\texttt{\_ZGVeNe8v\_sin}.

\subsection{SLEEF and the Vector Function ABI}

SLEEF is provided as two separate libraries. The first library exposes
the functions of SLEEF to programmers for inclusion in their C/C++
code. The second library exposes the functions with names mangled
according to the Vector Function ABI. This makes SLEEF a viable
alternative to libm and its SIMD counterpart \texttt{libmvec}, in
glibc. This also enables a user work-flow that relies on the
auto-vectorization capabilities of a compiler. The compatibility with
libmvec enables users to swap from libmvec to libsleef by simply
changing compiler options, without changing the code that generated
the vector call. The two SLEEF libraries are built from the same
source code, which are configured to target the different versions via
auto-generative programs that transparently rename the functions
according to the rules of the target library.

\section{Overview of Library Implementation}

\label{sec:overview}

One of the objectives of the SLEEF project is to provide a library of
vectorized math functions that can be used in conjunction with
vectorizing compilers. When a non-vectorized code is automatically
vectorized, the compiler converts calls to scalar math functions to calls to a
SIMD version of the math functions. In order to make
this conversion safe and applicable to wide variety of codes, we need
functions with 1-ULP error bound that conforms to ANSI C standard. On
the other hand, there are users who need better performance. Our
library provides 1-ULP accuracy version and 3.5-ULP accuracy version
for most of the functions. We confirmed that our library satisfies the accuracy requirements
on an empirical basis. For non-finite inputs and outputs, we implemented the
functions to return the same results as libm, as
specified in the ANSI C standard. They do not set errno nor raise an
exception.

In order to optimize a program with SIMD instructions, it is important
to eliminate conditional branches as much as possible, and execute the
same sequence of instructions regardless of the argument. If the algorithm requires
conditional branches according to the argument, it must prepare for
the case where the elements in the input vector contain both values
that would make a branch happen and not happen. 
Recent processors have a long pipeline and therefore branch
misprediction penalty can reach more than 10 cycles{\cite{1620789}}. Making a decision
for a conditional branch also requires non-negligible computation, within the scope of our tests.
A conditional move is an operator for choosing one value from two given
values according to a condition. This is equivalent to a ternary
operator and can be used in a vectorized code to replace a conditional
branch. Some other operations are also expensive in vectorized
implementation. A table-lookup is expensive. Although in-register
table lookup is reported fast on Cell BE SPU~\cite{4731241}, it is
substantially slower than polynomial evaluation without any table
lookup, within the scope of our tests. Most vector extensions do not provide
64-bit integer multiplication or a vector shift operator with which
each element of a vector can be specified a different number of bits
to shift. On the other hand, FMA and round-to-integer instructions are
supported by most vector extensions. {\changed Due to the nature of
  the evaluation methods, dependency between operations cannot be
  completely eliminated. Latencies of operations become an issue when
  a series of dependent operations are executed.}
FP division and square root are not too {\changed expensive from this
  aspect.}\footnote{The latencies of 256-bit DP add, divide and sqrt
  instructions are 4, 14 and 18 cycles, respectively on Intel Skylake
  processors~\cite{inteloptimize}.}

The actual structure of the pipeline in a processor is complex,
and such level of details are not well-documented for most CPUs. Therefore, it is not easy to
optimize the code according to such hardware implementation.  In this
paper, we define the latency and throughput of an instruction or a
subroutine as follows~\cite{Agner}. The latency of an instruction or a
subroutine is the delay that it generates in a dependency chain.  The
throughput is the maximum number of instructions or subroutines of the
same kind that can be executed per unit time when the inputs are
independent of the preceding instructions or subroutines. Several
tools and methods are proposed for automatically constructing models of
latency, throughput, and port usage of
instructions~\cite{Abel:2019:UCL:3297858.3304062, llvm-exegesis}. Within the scope of our tests, most
of the instruction latency in the critical path of evaluating a vector
math function tends to be dominated by FMA operations.
In many processors, FMA units are implemented in a
pipeline manner. Some powerful processors have multiple FMA units with
out-of-order execution, and thus the throughput of FMA instruction is
large, while the latency is long. In SLEEF, we try to maximize the
throughput of computation in a versatile way by only taking account of
dependencies among FMA operations. We regard each FMA operation as a
job that can be executed in parallel and try to reduce the length of
the critical path.

In order to evaluate a double-precision (DP) function to 1-ULP accuracy,
the internal computation with accuracy better than 1 ULP is sometimes
required. \emph{Double-double (DD)} arithmetic, in which a single
value is expressed by a sum of two double-precision FP values
~\cite{Dekker:1971:FTE:2716631.2717032,Shewchuk96adaptiveprecision},
is used for this purpose. All the basic operators for DD arithmetic
can be implemented without a conditional branch, and therefore it is
suitable for vectorized implementation. Because we only need 1-ULP
overall accuracy for DP functions, we use simplified DD
operators with less than the full DD accuracy. In SLEEF,
we omit re-normalization of DD values by default, allowing overlap
between the two numbers. We carry out re-normalization only when
necessary.

Evaluation of an elementary function often consists of three steps:
range reduction, approximation, and
reconstruction~\cite{Muller:1997:EFA:261217}. An approximation step
computes the elementary function using a polynomial. Since this
approximation is only valid for a small domain, a number within that
range is computed from the argument in a range reduction step. The
reconstruction step combines the results of the first two steps to
obtain the resulting number. 

An argument reduction method that finds an FP remainder of dividing
the argument $x$ by $\pi$ is used in evaluation of trigonometric
functions. The range reduction method suggested by Cody and
Waite~\cite{CodyWaite, Cody:1982:ITF} is used for small arguments.
The Payne and Hanek's method~\cite{Payne:1983:RRT:1057600.1057602}
provides an accurate range-reduction for a large argument of
trigonometric function, but it is expensive in terms of operations.

There are tools available for generating the coefficients of the
polynomials, such as Maple~\cite{10.1007/3-540-12868-9_95} and
Sollya~\cite{ChevillardJoldesLauter2010}. In order to fine-tune the
generated coefficients, we created a tool for generating coefficients
that minimizes the maximum relative error. When a SLEEF function evaluates a polynomial, it evaluates
a few lowest degree terms in DD precision
while other terms are computed in double-precision, in order to
achieve 1-ULP overall accuracy. Accordingly, coefficients in DD
precision or coefficients that can be represented by FP numbers with a
few most significant bits in mantissa are used in the last few terms.
We designed our tool to generate such
coefficients. We use Estrin's
scheme~\cite{Estrin:1960:OCS:1460361.1460365} to evaluate a polynomial
to reduce dependency between FMA operations.
This scheme reduces bubbles in the pipeline, and allows more
FMA operations to be executed in parallel. Reducing latency can improve the throughput of evaluating a function because
the latency and the reciprocal throughput of the entire function are
close to each other.


Below, we describe and compare the implementations of selected
functions in SLEEF, FDLIBM~\cite{FDLIBM} and Christoph Lauter's
Vector-libm~\cite{7869070}. We describe 1-ULP accuracy version of
functions in SLEEF. The error bound specification of FDLIBM is 1 ULP.

\subsection{Implementation of {\tt sin} and {\tt cos}}

FDLIBM uses Cody-Waite range reduction if the argument is under
{$2^{18}\pi$}. Otherwise, it uses the Payne-Hanek range reduction. Then, it
switches between polynomial approximations of the sine and cosine
functions on $[-\pi/4, \pi/4]$. Each polynomial has 6 non-zero terms.

{\tt sin} and {\tt cos} in Vector-libm have 4-ULP error bound. They
use a vectorized path if all arguments are greater
than 3.05e-151, and less than 5.147 for sine and 2.574 for cosine. In
the vectorized paths, a polynomial with 8 and 9 non-zero terms is used
to approximate the sine function on $[-\pi/2, \pi/2]$, following
Cody-Waite range reduction. In the scalar paths, Vector-libm uses a polynomial with 10
non-zero terms.

SLEEF switches among two Cody-Waite range reduction methods with
approximation with different sets of constants, and the Payne-Hanek
reduction. The first version of the algorithm operates for arguments
within $[-15, 15]$, and the second version for arguments that are within
$[-10^{14}, 10^{14}]$.
Otherwise, SLEEF uses a vectorized Payne-Hanek reduction,
which is described in \ref{sec:paynehanek}. 
SLEEF only uses conditional branches for choosing a reduction
method from Cody-Waite and Payne-Hanek. SLEEF uses a polynomial
approximation of the sine function on $[-\pi/2, \pi/2]$, which has 9
non-zero terms. The sign is set in the reconstruction step.

\subsection{Implementation of {\tt tan}}

\label{sec:tan}

After Cody-Waite or Payne-Hanek reduction, FDLIBM reduces the argument
to $[0,0.67434]$, and uses a polynomial approximation with 13 non-zero
terms. It has 10 {\tt if} statements after Cody-Waite reduction.

{\tt tan} in Vector-libm has 8-ULP error bound. A vectorized path is
used if all arguments are less than 2.574 and greater
than 3.05e-151. After Cody-Waite range reduction, a polynomial with 9
non-zero terms for approximating sine function on $[-\pi/2, \pi/2]$ is
used twice to approximate sine and cosine of the reduced argument. The
result is obtained by dividing these values. In the scalar path, Vector-libm evaluates a
polynomial with 10 non-zero terms twice.

In SLEEF, the argument is reduced in 3 levels. It first reduces the argument
to $[-\pi/2, \pi/2]$ with Cody-Waite or Payne-Hanek range
reduction. Then, it reduces the argument to $[-\pi/4, \pi/4]$ with
$\tan a_1 = 1 / \tan(\pi/2 - a_0)$. At the third level, it reduces the argument
to $[-\pi/8, \pi/8]$ with the double-angle formula. Let
$a_0$ be the reduced argument with Cody-Waite or Payne-Hanek. $a_1 =
\pi/2 - a_0$ if $|a_0| > \pi/4$. Otherwise, $a_1 = a_0$. Then, SLEEF uses a
polynomial approximation of the tangent function on $[-\pi/8, \pi/8]$,
which has 9 non-zero terms, to approximate $\tan (a_1/2)$. Let
$t$ be the obtained value with this approximation. Then, $\tan a_0
\approx 2t / (1 - t^2)$ if $|a_0| \leq \pi/4$. Otherwise, $\tan a_0
\approx (1 - t^2) / (2t)$. SLEEF only uses conditional branches for
choosing a reduction method from Cody-Waite and Payne-Hanek. Annotated
source code of {\tt tan} is shown in Appendix~\ref{sec:sourcecode}

\subsection{Implementation of {\tt asin} and {\tt acos}}

FDLIBM and SLEEF first reduces the argument to $[0, 0.5]$ using
$\arcsin x = \pi/2-2\arcsin \sqrt{(1-x)/2}$ and $\arccos x = 2\arcsin
\sqrt{(1-x)/2}$.

Then, SLEEF uses a polynomial approximation of arcsine on $[0, 0.5]$
with 12 non-zero terms.

FDLIBM uses a rational approximation with 11 terms (plus one
division). For computing arcsine, FDLIBM switches the approximation
method if the original argument is over 0.975. For computing
arccosine, it has three paths that are taken when $|x|<0.5$, $x \leq
-0.5$ and $x \geq 0.5$, respectively. It has 7 and 6 {\tt if}
statements in {\tt asin} and {\tt acos}, respectively.

{\tt asin} and {\tt acos} in Vector-libm have 6-ULP error bound.
{{\tt asin} and {\tt acos} in Vector-libm use} vectorized paths if arguments are all
greater than 3.05e-151 and 2.77e-17, respectively. Vector-libm evaluates polynomials with 3,
8, 8, and 5 terms to compute arcsine. It evaluates a polynomial with
21 terms for arccosine.

\subsection{Implementation of {\tt atan}}

FDLIBM reduces the argument to $[0, 7/16]$. It uses a
polynomial approximation of the arctangent function with 11 non-zero
terms. It has 9 {\tt if} statements.

{\tt atan} in Vector-libm have 6-ULP error bound. Vector-libm uses vectorized paths
if arguments are all greater than 1.86e-151
and less than 2853. It evaluates four polynomials with 7, 9, 9 and 4 terms
in the vectorized path.

SLEEF reduces argument $a$ to $[0, 1]$ using $\arctan x = \pi/2 -
\arctan(1/x)$. Let $a' = 1/a$ if $|a| \geq 1$. Otherwise, $a' = a$. It
then uses a polynomial approximation of arctangent function with 20
non-zero terms to approximate $r \approx \arctan a'$.  As a
reconstruction, it computes $\arctan a \approx \pi/2 - r$ if $|a| \geq
1$. Otherwise, $\arctan a \approx r$.

\subsection{Implementation of {\tt log}}

FDLIBM reduces the argument to $[\sqrt{2}/2, \sqrt{2}]$. It then
approximates the reduced argument with a polynomial that contains 7 non-zero
terms in a similar way to SLEEF. It has 9 {\tt if}
statements.

{\tt log} in Vector-libm has 4-ULP error bound. It uses a vectorized path
if the input is a normalized number. It uses a polynomial with 20
non-zero terms to approximate the logarithm function on
$[0.75, 1.5]$. It does not use division.

SLEEF multiplies the argument $a$ by $2^{64}$, if the argument
is a denormal number. Let $a'$ be the resulting argument, $e = \lfloor
\log_2 (4a'/3) \rfloor$ and $m = a' \cdot 2^{-e}$. If $a$ is a
denormal number, $e$ is subtracted 64. SLEEF uses a polynomial with 7 non-zero
terms to evaluate $\log m \approx \sum_{n=0}^{6} C_n \left(
\frac{m-1}{m+1} \right)^{2n+1}$, where $C_0 ... C_6$ are constants. As
a reconstruction, it computes $\log a = e\log 2 + \log m$.

\subsection{Implementation of {\tt exp}}

All libraries reduce the argument range to $[-(\log 2) / 2, (\log 2) /
  2]$ by finding $r$ and integer $k$ such that $x = k \log 2 + r, |r|
\leq (\log2) /2$.

SLEEF then uses a polynomial approximation with 13 non-zero terms to
directly approximate the exponential function of this domain.
It achieves 1-ULP error bound without using a DD operation.

FDLIBM uses a polynomial with 5 non-zero terms to approximate $f(r) =
r(e^r+1)/(e^r-1)$. It then computes {\tt exp}$(r) = 1 +
2r/(f(r)-r)$. It has 11 {\tt if} statements.

The reconstruction step is to add integer $k$ to the exponent of the
resulting FP number of the above computation.

{\tt exp} in Vector-libm has 4-ULP error bound. A vectorized path
covers almost all input domains. It uses a polynomial with 11 terms to
approximate the exponential function.

\subsection{Implementation of {\tt pow}}

FDLIBM computes $y \log_2x$ in DD precision.  Then, it computes {\tt
  pow}($x$, $y$) = $e^{\log 2 \cdot y \log_2x}$. It has 44 {\tt if}
statements.

Vector-libm does not implement {\tt pow}.

SLEEF computes $e^{y \log x}$. The internal computation is carried out
in DD precision. In order to compute logarithm internally, it uses a
polynomial with 11 non-zero terms. The accuracy of the internal
logarithm function is around 0.008 ULP. The internal exponential
function in {\tt pow} uses a polynomial with 13 non-zero terms.

\subsection{The Payne-Hanek Range Reduction}

\label{sec:paynehanek}

Our method computes $\mathrm{rfrac}(2x/\pi) \cdot \pi/2$, where
$\mathrm{rfrac}(a) \coloneqq a - \mathrm{round}(a)$. The argument $x$
is an FP number, and therefore it can be represented as $M \cdot 2^E$,
where $M$ is an integer mantissa and $E$ is an integer exponent
$E$. We now denote the integral part and the fractional part of $2^E
\cdot 2/\pi$ as $I(E)$ and $F(E)$, respectively. Then,
\begin{align*}
\mathrm{rfrac}(2x/\pi) &= \mathrm{rfrac}(M \cdot 2^E \cdot 2/\pi) \\
&= \mathrm{rfrac}(M \cdot (I(E) + F(E))) \\
&= \mathrm{rfrac}(M \cdot F(E)).
\end{align*}

The value $F(E)$ only depends on the exponent of the argument, and
therefore, can be calculated and stored in a table, in advance.  In
order to compute $\mathrm{rfrac}(M \cdot F(E))$ in DD precision,
$F(E)$ must be in quad-double-precision. We now denote $F(E) = F_0(E)
+ F_1(E) + F_2(E) + F_3(E)$, where $F_0(E) ... F_3(E)$ are DP numbers
and $|F_0(E)| \geq |F_1(E)| \geq |F_2(E)| \geq |F_3(E)|$. Then,
\begin{align}
  & \mathrm{rfrac}(M \cdot F(E)) \nonumber \\
= & \mathrm{rfrac}(M \cdot F_0(E) + M \cdot F_1(E) \nonumber \\
  &  + M \cdot F_2(E) + M \cdot F_3(E)) \nonumber \\
= & \mathrm{rfrac}(\mathrm{rfrac}(\mathrm{rfrac}(\mathrm{rfrac}(M \cdot F_0(E)) + M \cdot F_1(E)) \nonumber \\
  &  + M \cdot F_2(E)) + M \cdot F_3(E)), \label{eq:ph0}
\end{align}
\noindent
because $\mathrm{rfrac}(a + b) = \mathrm{rfrac}(\mathrm{rfrac}(a) +
b)$. In the method, we compute (\ref{eq:ph0}) in DD precision in order
to avoid overflow. The size of the table retaining $F_0(E) ... F_3(E)$
is 32K bytes. Our method is included in the source code of {\tt tan}
shown in Appendix~\ref{sec:sourcecode}.

\begin{figure*}[tb]
\centering
\vspace{3mm}
\includegraphics[width=0.9\textwidth]{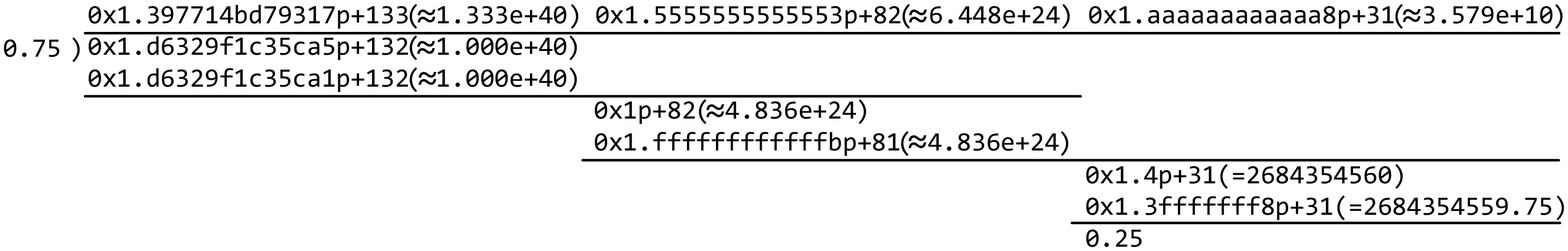}
\caption{Example computation of FP remainder}
\label{fig:fmod}
\end{figure*}

FDLIBM seems to implement the original Payne-Hanek algorithm with more
than 100 lines of C code, which includes 13 {\tt if} statements, 18
{\tt for} loops, 1 {\tt switch} statement and 1 {\tt goto}
statement. The numbers of iterations of most of the {\tt for} loops
depend on the argument.

Vector-libm implements a non-vectorized variation of the Payne-Hanek algorithm which has
some similarity with our method. In order to reduce argument $x$, it
first decomposes $|x|$ into $E$ and $n$ such that $2^E \cdot n =
|x|$. A triple-double (TD) approximation to $t(E) = 2^E / \pi - 2 \cdot
\lfloor 2^{E - 1} / \pi \rfloor$ is looked-up from a table. It then
calculates $m = n \cdot t(E)$ in TD. The reduced argument is obtained
as a product of $\pi$ and the fractional part of $m$. In Table~\ref{table:paynehanek},
we compare the numbers of FP operators in the
implementations. Note that the method in Vector-libm is used for
trigonometric functions with 4-ULP error bound, while our method is
used for functions with 1-ULP error bound.

\begin{table}[t]
\begin{center}
\caption{Number of FP operators in the Payne-Hanek implementations}
\label{table:paynehanek}
\begin{tabular}{c|c|c}
\hline
Operator & SLEEF (1-ULP) & Vector-libm (4-ULP) \\
\hline
add/sub & 36 & 71\\
mul     &  5 & 18\\
FMA     & 11 &  0\\
round   &  8 &  0\\
\hline
\end{tabular}
\end{center}
\end{table}

\subsection{FP Remainder}

\label{sec:remainder}

We devised an exact remainder calculation method suitable for
vectorized implementation. The method is based on the long division
method, where an FP number is regarded as a digit. Fig.~\ref{fig:fmod}
shows an example process for calculating the FP remainder of 1e+40 /
0.75. Like a {\changed typical} long division, we first find integer quotient
1.333e+40 so that 1.333e+40 $\cdot 0.75$ does not exceed 1e+40.  We
multiply the found quotient with $0.75$, and then subtract it from
1e+40 to find the dividend 4.836e+24 for the second iteration.

Our basic algorithm is shown in Algorithm~\ref{algorithm:fmod}. If
$n$, $d$ and $q_k$ are FP numbers of the same precision $p$, then
$r_k$ is representable with an FP number of precision $2p$. In this
case, the number of iterations can be minimized by substituting $q_k$
with the largest FP number of precision $p$ within the range specified
at line~\ref{line:q_k}. However, the algorithm still works if $q_k$ is
any FP number of precision $p$ within the range. By utilizing this
property, an implementer can use a division operator that does not
return a correctly rounded result. The source code of an
implementation of this algorithm is shown in Fig.~\ref{fig:xfmod} in
Appendix~\ref{sec:sourcecodexfmod}. A part of the proof of correctness is
shown in Appendix~\ref{sec:prooffmod}.

FDLIBM uses a method of shift and subtract. It first converts the
mantissa of two given arguments into 64-bit integers, and calculates a
remainder in a bit-by-bit basis. The main loop iterates $ix-iy$ times,
where $ix$ and $iy$ are the exponents of the arguments of \texttt{fmod}. This loop includes
10 integer additions and 3 {\tt if} statements. The number of
iterations of the main loop can reach more than 1000.

Vector-libm does not implement FP remainder.

\begin{algorithm}[tb]
\caption{Exact remainder calculation}
\label{algorithm:fmod}
{\fontsize{8}{10}\selectfont
\begin{algorithmic}[1]                
\REQUIRE Finite positive numbers $n$ and $d$
\ENSURE Returns $n - d\lfloor n/d \rfloor$

\STATE $r_0 := n, k := 0$
\WHILE{$d \leq r_k$}
\STATE $q_k$ is an arbitrary integer satisfying $(r_k/d)/2 \leq q_k \leq r_k/d$ \label{line:q_k}
\STATE $r_{k+1} := r_k - q_kd$
\STATE $k := k + 1$
\ENDWHILE
\RETURN $r_k$
\end{algorithmic}
}
\end{algorithm}

\subsection{Handling of Special Numbers, Exception and Flags}

Our implementation gives a value within the specified error bound
without special handling of denormal numbers, unless otherwise noted.

When a function has to return a specific value for a specific value of an argument
(such as a NaN or a negative zero) is given, such a condition is checked
at the end of each function. The return value is substituted with the
special value if the condition is met. This process is complicated in
functions like {\tt pow}, because they have many conditions for
returning special values.

SLEEF functions do not give correct results if the computation mode is
different from round-to-nearest. They do not set errno nor raise an
exception. This is a common behavior among vectorized math libraries
including Libmvec~\cite{libmvec} and SVML~\cite{SVML}. Because of SIMD
processing, functions can raise spurious exceptions if they try to
raise an exception.

\subsection{Summary}

FDLIBM extensively uses conditional branches in order to
switch the polynomial according to the argument({\tt sin}, {\tt cos}, {\tt tan}, {\tt log},
etc), to return a special value if the arguments are special
values({\tt pow}, etc.), and to control the number of
iterations (the Payne-Hanek reduction).

Vector-libm switches between a few polynomials in most of the
functions. It does not provide functions with 1-ULP error bound, nevertheless,
the numbers of non-zero terms in the polynomials are
larger than other two libraries in some of the functions. A
vectorized path is used only if the argument is smaller than 2.574 in
{\tt cos} and {\tt tan}, although these functions are frequently evaluated with an argument up to $2\pi$.  In
most of the functions, Vector-libm uses a non-vectorized path if the
argument is very small or a non-finite number. For example, it processes 0 with
non-vectorized paths in many functions, although 0 is a
frequently evaluated argument in normal situations. If non-finite
numbers are once contained in data being processed, the whole
processing can become significantly slower afterward. Variation in
execution time can be exploited for a side-channel attack in
cryptographic applications.

SLEEF uses the fastest paths if all the arguments are under 15 for
trigonometric functions, and the same vectorized path is used
regardless of the argument in most of the non-trigonometric
functions. SLEEF always uses the same polynomial regardless of the
argument in all functions.

Although reducing the number of conditional branches has a few
advantages in implementing vector math libraries, it seems to be not
given a high priority in other libraries.

\section{Testing}

\label{sec:testing}

SLEEF includes three kinds of testers. The first two kinds of testers
test the accuracy of all functions against high-precision evaluation
using the MPFR library. In these tests, the computation error in ULP
is calculated by comparing the values output by each SLEEF function
and the values output by the corresponding function in the MPFR library,
and it is checked if the error is within the specified bounds.

\subsection{Perfunctory Test}

The first kind of tester carries out a perfunctory set of tests to
check if the build is correct. These tests include standards
compliance tests, accuracy tests and regression tests.

In the standards compliance tests, we test if the functions return the
correct values when values that require special handling are given as
the argument. These argument values include $\pm$Inf, NaN and
$\pm$0. Atan2 and pow are binary functions and have many
combinations of these special argument values. These are also all tested.

In the accuracy test, we test if the error of the returned values from the
functions is within the specified range, when a predefined set of
argument values are given. These argument values are basically chosen
between a few combinations of two values at regular intervals. The
trigonometric functions are also tested against argument values close
to integral multiples of $\pi/2$. Each function is tested against tens of thousands
of argument values in total.

In the regression test, the functions are tested with argument values
that triggered bugs in the previous library release, in order to
prevent re-emergence of the same bug.

The
executables are separated into a tester and IUTs (
Implementation Under Test). The tests are carried out by making these
two executables communicate via an input/output pipeline, in order to enable testing of
libraries for architectures which the MPFR library does not
support.

\subsection{Randomized Test}

The second kind of tester is designed to run continuously. This tester
generates random arguments and compare the output from each function
to the output calculated with the corresponding function in the MPFR
library. This tester is expected to find bugs if it is run for
a sufficiently long time.

In order to randomly generate an argument, the tester generates random
bits of the size of an FP value, and reinterprets the bits as an FP
value. The tester executes the randomized
test for all the functions in the library at several thousand
arguments per second for each function on a computer with a Core
i7-6700 CPU.

In the SLEEF project, we use randomized testing in order to check the
correctness of functions, rather than formal verification. It is
indeed true that proving correctness of implementation contributes to
the reliability of implementation. However, there is a performance
overhead because the way of implementation is limited in a form that is
easy to prove the correctness. There would be an increased cost of
maintaining the library because of the need for updating the proof
each time the implementation is modified.

\subsection{Bit-Identity Test}

The third kind of tester is for testing if bit-identical results are
returned from the functions that are supposed to return such results.
This test is designed to compare the results among the binaries
compiled with different vector extensions. For each predetermined list of
arguments, we calculate an MD5 hash value of all the outputs from each
function. Then, we check if the hash values match among functions for
different architectures.

\section{Performance Comparison}

\label{sec:performance}

In this section, we present results of a performance comparison
between FDLIBM Version 5.3~\cite{FDLIBM}, Vector-libm~\cite{7869070},
SLEEF 3.4, and Intel SVML~\cite{SVML} included in Intel C Compiler 19.

We measured the reciprocal throughput of each function by
measuring the execution time of a tight loop that repeatedly calls the
function in a single-threaded process. In order to obtain useful
results, we turned off optimization flags when compiling the source code of
this tight loop,\footnote{\changed
If we turn on the optimizer, there is concern that the compiler
optimizes away the call to a function. In order to prevent this, we
have to introduce extra operations, but this also introduces
overhead. After trying several configurations of the loop and
optimizer settings, we decided to turn off the optimizer in favor of
reproducibility, simplicity and fairness. We checked the assembly
output from the compiler and confirmed that the unoptimized loop
simply calls the target function and increments a counter, and
therefore that the operations inside a loop are minimal.
} while the libraries are compiled with their default
optimization options. We did not use LTO. We confirmed that the calls
to the function are not compiled out or inlined by checking the
assembly output from the compiler. The number of function calls by
each loop is $10^{10}$, and the execution time of this loop is
measured with the {\tt clock\_gettime} function.

We compiled SLEEF and FDLIBM using gcc-7.3.0 with ``{\tt -O3 -mavx2
  -mfma}'' optimization options.
We compiled Vector-libm using gcc-7.3.0 with the default ``{\tt -O3
  -march=native -ftree-vectorize -ftree-vectorizer-verbose=1
  -fno-math-errno}'' options. We changed VECTOR\_LENGTH in {\tt
  vector.h} to 4 and compiled the source code on a computer with an
Intel Core i7-6700 CPU.\footnote{Vector-libm evaluates functions with 512
  bits of vector length by default. Because SLEEF and SVML are
  256-bit wide, the setting is changed.}
The accuracy of functions in SVML can be chosen by compiler
options. We specified
an ``{\tt -fimf-max-error=1.0}'' and
an ``{\tt -fimf-max-error=4.0}'' options for icc to
obtain the 1-ULP and 4-ULP accuracy results, respectively.

We carried out all the measurements on a physical PC with Intel Core i7-6700
CPU @ 3.40GHz without any virtual machine. In order to make sure that the CPU is always running at
the same 3.4GHz clock speed during the measurements, we turned off
Turbo Boost. With this setting, 10 nano sec. corresponds to 34 clock
cycles.

The following results compare the the reciprocal throughput of each
function. If the implementation is vectorized and each vector has $N$
elements of FP numbers, then a single execution evaluates the
corresponding mathematical function $N$ times.
We generated arguments in advance and stored in arrays. Each time a
function is executed, we set a randomly generated argument to each
element of the argument vector (each element is set with a different
value). The measurement results do not include the delay for generating
random numbers.

\subsection{Execution Time of Floating Point Remainder}

\begin{figure}[tb]
\centering
\includegraphics[width=0.7\columnwidth]{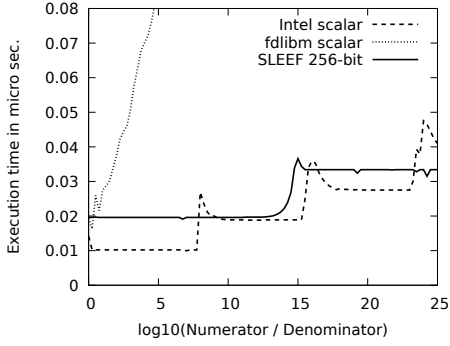}
\caption{Reciprocal throughput of double-precision fmod functions}
\label{fig:benchfmod}
\end{figure}

We compared the reciprocal throughput of double-precision {\tt fmod} functions in
the \texttt{libm} included in Intel C Compiler 19, FDLIBM and
SLEEF. All the FP remainder functions always return a
correctly-rounded result. We generated a random denominator $d$ and a
numerator uniformly distributed within $[1, 100]$ and $[0.95r\cdot d,
  1.05r\cdot d]$, respectively, where $r$ is varied from 1 to
$10^{25}$. Fig.~\ref{fig:benchfmod} shows the reciprocal throughput of the {\tt
  fmod} function in each library. Please note that SVML does not
contain a vectorized {\tt fmod} function.

The graph of reciprocal throughput looks like a step function, 
because the number of iterations increases in this way.

\subsection{Comparison of Overall Execution Time}

\begin{table}[tb]
\begin{center}
\caption{Reciprocal throughput in nano sec.}
\label{table:overall}
\begin{tabular}{c|c|c|c|c}
\hline
Func, error bound, & \multirow{2}{*}{Vector-libm} & \multirow{2}{*}{FDLIBM} & \multirow{2}{*}{SLEEF} & \multirow{2}{*}{SVML} \\
domain&&&&\\
\hline \hline
sin, 1 ulp, & \multirow{2}{*}{} & \multirow{2}{*}{4.927} & \multirow{2}{*}{11.43} & \multirow{2}{*}{13.68} \\
$[0.4, 0.5]$&&&&\\
\hline
sin, 4 ulps, & \multirow{2}{*}{9.601} & \multirow{2}{*}{} & \multirow{2}{*}{7.504} & \multirow{2}{*}{6.679} \\
$[0.4, 0.5]$&&&&\\
\hline
sin, 1 ulp, & \multirow{2}{*}{} & \multirow{2}{*}{18.96} & \multirow{2}{*}{11.41} & \multirow{2}{*}{13.86} \\
$[0, 6.28]$&&&&\\
\hline
sin, 4 ulps, & \multirow{2}{*}{12.48} & \multirow{2}{*}{} & \multirow{2}{*}{7.507} & \multirow{2}{*}{6.723} \\
$[0, 6.28]$&&&&\\
\hline
sin, 1 ulp, & \multirow{2}{*}{} & \multirow{2}{*}{162.3} & \multirow{2}{*}{48.79} & \multirow{2}{*}{41.72} \\
$[0, 1e+100]$&&&&\\
\hline
sin, 4 ulps, & \multirow{2}{*}{288.6} & \multirow{2}{*}{} & \multirow{2}{*}{43.82} & \multirow{2}{*}{34.96} \\
$[0, 1e+100]$&&&&\\
\hline
cos, 1 ulp, & \multirow{2}{*}{} & \multirow{2}{*}{11.42} & \multirow{2}{*}{13.75} & \multirow{2}{*}{12.99} \\
$[0.4, 0.5]$&&&&\\
\hline
cos, 4 ulps, & \multirow{2}{*}{9.557} & \multirow{2}{*}{} & \multirow{2}{*}{7.850} & \multirow{2}{*}{7.917} \\
$[0.4, 0.5]$&&&&\\
\hline
cos, 1 ulp, & \multirow{2}{*}{} & \multirow{2}{*}{18.45} & \multirow{2}{*}{13.74} & \multirow{2}{*}{13.18} \\
$[0, 6.28]$&&&&\\
\hline
cos, 4 ulps, & \multirow{2}{*}{13.97} & \multirow{2}{*}{} & \multirow{2}{*}{7.850} & \multirow{2}{*}{7.838} \\
$[0, 6.28]$&&&&\\
\hline
cos, 1 ulp, & \multirow{2}{*}{} & \multirow{2}{*}{162.1} & \multirow{2}{*}{50.38} & \multirow{2}{*}{38.38} \\
$[0, 1e+100]$&&&&\\
\hline
cos, 4 ulps, & \multirow{2}{*}{360.3} & \multirow{2}{*}{} & \multirow{2}{*}{46.09} & \multirow{2}{*}{36.57} \\
$[0, 1e+100]$&&&&\\
\hline
tan, 1 ulp, & \multirow{2}{*}{} & \multirow{2}{*}{7.819} & \multirow{2}{*}{17.30} & \multirow{2}{*}{15.71} \\
$[0.4, 0.5]$&&&&\\
\hline
tan, 4+ ulps, & \multirow{2}{*}{15.58} & \multirow{2}{*}{} & \multirow{2}{*}{9.367} & \multirow{2}{*}{7.570} \\
$[0.4, 0.5]$&&&&\\
\hline
tan, 1 ulp, & \multirow{2}{*}{} & \multirow{2}{*}{22.24} & \multirow{2}{*}{17.28} & \multirow{2}{*}{15.78} \\
$[0, 6.28]$&&&&\\
\hline
tan, 4+ ulps, & \multirow{2}{*}{20.16} & \multirow{2}{*}{} & \multirow{2}{*}{9.367} & \multirow{2}{*}{7.595} \\
$[0, 6.28]$&&&&\\
\hline
tan, 1 ulp, & \multirow{2}{*}{} & \multirow{2}{*}{177.0} & \multirow{2}{*}{48.82} & \multirow{2}{*}{43.31} \\
$[0, 1e+100]$&&&&\\
\hline
tan, 4+ ulps, & \multirow{2}{*}{399.4} & \multirow{2}{*}{} & \multirow{2}{*}{36.54} & \multirow{2}{*}{40.50} \\
$[0, 1e+100]$&&&&\\
\hline
asin, 1 ulp, & \multirow{2}{*}{} & \multirow{2}{*}{14.87} & \multirow{2}{*}{12.99} & \multirow{2}{*}{12.10} \\
$[-1, 1]$&&&&\\
\hline
asin, 4+ ulps, & \multirow{2}{*}{20.75} & \multirow{2}{*}{} & \multirow{2}{*}{5.552} & \multirow{2}{*}{9.627} \\
$[-1, 1]$&&&&\\
\hline
acos, 1 ulp, & \multirow{2}{*}{} & \multirow{2}{*}{12.07} & \multirow{2}{*}{16.09} & \multirow{2}{*}{12.11} \\
$[-1, 1]$&&&&\\
\hline
acos, 4+ ulps, & \multirow{2}{*}{23.62} & \multirow{2}{*}{} & \multirow{2}{*}{7.572} & \multirow{2}{*}{10.23} \\
$[-1, 1]$&&&&\\
\hline
atan, 1 ulp, & \multirow{2}{*}{} & \multirow{2}{*}{10.16} & \multirow{2}{*}{22.12} & \multirow{2}{*}{19.97} \\
$[-700, 700]$&&&&\\
\hline
atan, 4+ ulps, & \multirow{2}{*}{35.54} & \multirow{2}{*}{} & \multirow{2}{*}{9.251} & \multirow{2}{*}{12.09} \\
$[-700, 700]$&&&&\\
\hline
log, 1 ulp, & \multirow{2}{*}{} & \multirow{2}{*}{31.66} & \multirow{2}{*}{15.46} & \multirow{2}{*}{12.05} \\
$[0, 1e+300]$&&&&\\
\hline
log, 4 ulps, & \multirow{2}{*}{39.64} & \multirow{2}{*}{} & \multirow{2}{*}{9.636} & \multirow{2}{*}{8.842} \\
$[0, 1e+300]$&&&&\\
\hline
exp, 1 ulp, & \multirow{2}{*}{} & \multirow{2}{*}{12.19} & \multirow{2}{*}{7.663} & \multirow{2}{*}{7.968} \\
$[-700, 700]$&&&&\\
\hline
exp, 4 ulps, & \multirow{2}{*}{17.35} & \multirow{2}{*}{} & \multirow{2}{*}{} & \multirow{2}{*}{6.756} \\
$[-700, 700]$&&&&\\
\hline
pow, 1 ulp, & \multirow{2}{*}{} & \multirow{2}{*}{69.40} & \multirow{2}{*}{55.53} & \multirow{2}{*}{75.18} \\
$[-30, 30] [-30, 30]$&&&&\\
\hline
\end{tabular}
\end{center}
\end{table}

We compared the reciprocal throughput of 256-bit wide vectorized
double-precision functions in Vector-libm, SLEEF and SVML, and scalar
functions in FDLIBM.  We generated random arguments that were uniformly distributed within the
indicated intervals for each function. In order to check execution
speed of fast paths in trigonometric functions, we measured the reciprocal throughput
with arguments within $[0.4, 0.5]$. The result is shown in
Table~\ref{table:overall}.

The reciprocal throughput of functions in SLEEF is comparable to that of SVML in all
cases. This is because the latency of FP operations is generally dominant in the
execution time of math functions. Because there are two levels of
scheduling mechanisms, which {\changed includes} the optimizer in a compiler and the
out-of-order execution hardware, there is small room for
{\changed making a difference to the throughput or latency.}

Execution speed of FDLIBM is not very slow despite many conditional
branches. This seems to be because of a smaller number of FP
operations, and faster execution speed of scalar instructions compared
to equivalent SIMD instructions.

Vector-libm is slow even if only the vectorized path is used. This
seems to be because Vector-libm evaluates polynomials with a large
number of terms. Auto-vectorizers are still developing, and the
compiled binary code might not be well optimized. When a slow path has
to be used, Vector-libm is even slower since a scalar evaluation has
to be carried out for each of the elements in the vector.

Vector-libm uses Horner's method to evaluate polynomials, which
involves long latency of chained FP operations. In FDLIBM, this
latency is reduced by splitting polynomials into even and odd terms,
which can be evaluated in parallel. SLEEF uses Estrin's scheme. In our
experiments, there was only a small difference between Estrin's scheme
and splitting polynomials into even and odd terms with respect to
execution speed.

\section{Conclusion}

\label{sec:conclusion}

In this paper, we showed that our SLEEF library shows performance
comparable to commercial libraries while maintaining good
portability. We have been continuously developing SLEEF since
2010.\footnote{\url{https://sleef.org/}}~\cite{Shibata2010} {We distribute} SLEEF
under the Boost Software License~\cite{BSL}, which is a
permissive open source license. We actively communicate with
developers of compilers and members of other projects in order to
understand the needs of real-world users. The Vector Function ABI is important
in developing vectorizing compilers. The functions that return
bit-identical results are added to our library to reflect requests
from our multiple partners. We thoroughly tested these functionalities,
and SLEEF is already adopted in multiple commercial products.

\appendices

\section{Annotated source code of {\tt tan}}

\label{sec:sourcecode}

\begin{figure}[tb]
\begin{center}
\begin{tabular}{c}
\begin{lstlisting}
double xtan(double d) {
  double u;
  double2 x = dd(0, 0), y;
  int q = 0;

  if (fabs(d) < 15) { /*\label{line:tan_branch}*/
    // Cody-Waite
    q = round(d * M_2_PI); /*\label{line:tan_round0}*/
    u = fma(q, -0x1.921fb54442d18p0, d);
    x = ddadd_d2_d_d(u, q * -0x1.1a62633145c07p-54);
  } else {
    // Payne-Hanek /* \label{line:ph_begin} */
    int ex = ilogb(d), M = ex > 700 ? -130 : -2;
    if (ex < 0) ex = 0; /*\label{line:tan_cmov0}*/
    u = ldexp(d, M);
    for(int i=0;i<4;i++) { /*\label{line:tan_for} \label{line:ph_forloopbegin} */
      y = ddmul_d2_d_d(u, tab[ex*4+i]);
      x = ddadd2_d2_d2_d2(x, y);
      double r = round(4*x.x); /*\label{line:tan_round1}*/
      q += (int32_t)((r - 4*round(x.x))); /*\label{line:tan_round2}*/
      x.x -= r * 0.25;
      x = ddnormalize_d2_d2(x);
    } /* \label{line:ph_forloopend} */
    x = ddmul_d2_d2_d2(x, /* \label{line:ph_mulpio2} */
      dd(0x1.921fb54442d18p2, 0x1.1a62633145c07p-52));
    if (!isfinite(d)) x.x = NAN; // NAN handling /*\label{line:tan_cmov1} \label{line:ph_end}*/
  }

  // Reduction with double-angle formula
  y = ddscale_d2_d2_d(x, 0.5);

  // Polynomial evaluation with Estrin's scheme
  // Domain : |y| <= PI/8
  x = ddsqu_d2_d2(y);
  u = ESTRIN(x.x, /*\label{line:tan_estrin}*/
	     +0.3245098826639276316e-3,
	     +0.5619219738114323735e-3,
	     +0.1460781502402784494e-2,
	     +0.3591611540792499519e-2,
	     +0.8863268409563113126e-2,
	     +0.2186948728185535498e-1,
	     +0.5396825399517272970e-1,
	     +0.1333333333330500581e+0);

  // Last two terms are evaluated with Horner's method
  u = fma(u, x.x, +0.3333333333333343695e+0);

  // Last term is evaluated in DD precision
  x = ddadd_d2_d2_d2(y,
        ddmul_d2_d2_d(ddmul_d2_d2_d2(x, y), u));

  // Reconstruction with double-angle formula
  y = ddadd_d2_d_d2(-1, ddsqu_d2_d2(x));
  x = ddscale_d2_d2_d(x, -2);

  // Reconstruction with tan(PI/2 - x) = cot(x)
  if (q & 1) { /*\label{line:tan_cmov2}*/
    double2 t = x; x = y; y.x = -t.x; y.y = -t.y;
  }
  x = dddiv_d2_d2_d2(x, y);

  if (d == 0) return d; // Negative-zero handling /*\label{line:tan_cmov3}*/
  return x.x + x.y;
}
\end{lstlisting}
\end{tabular}
\caption{C source code of {\tt tan}}
\label{fig:tan}
\end{center}
\end{figure}

\begin{figure}[tb]
\begin{center}
\begin{tabular}{c}
\begin{lstlisting}
#include <mpfr.h>

double tab[4096];

void init() {
  mpfr_set_default_prec(1280);

  mpfr_t pi, twoopi, m;
  mpfr_inits(pi, twoopi, m, NULL);
  mpfr_const_pi(pi, GMP_RNDN);
  mpfr_d_div(twoopi, 2, pi, GMP_RNDN);

  for(int ex=0;ex<1024;ex++) {
    int M = ex > 700 ? -128 : 0;
    mpfr_set(m, twoopi, GMP_RNDN);
    mpfr_set_exp(m, mpfr_get_exp(m) + (ex - 53));
    mpfr_frac(m, m, GMP_RNDN);
    mpfr_set_exp(m, mpfr_get_exp(m) - (ex - 53 + M));

    for(int i=0;i<4;i++) {
      union { double d; int64_t i; } tmp = { .d = mpfr_get_d(m, GMP_RNDN) };
      tmp.i &= 0xfffffffffffffffeLL;
      tab[ex*4+i] = tmp.d;
      mpfr_sub_d(m, m, tab[ex*4+i], GMP_RNDN);
    }
  }

  mpfr_clears(pi, twoopi, m, NULL);
}
\end{lstlisting}
\end{tabular}
\caption{C source code of Payne-Hanek table generation}
\label{fig:gentab}
\end{center}
\end{figure}

Fig.~\ref{fig:tan} and \ref{fig:gentab} shows a C source code of our implementation of the
tangent function with 1-ULP error bound. We omitted the second Cody-Waite reduction
for the sake of simplicity. The definitions of DD operators
are provided in Table~\ref{table:ddfunc}. In the implementation of our
library, we wrote all the operators with VEAL, as described in
Sec.~\ref{sec:veal}. The only conditional branch in this source code
is the {\tt if} statement at line~\ref{line:tan_branch}. We implemented the other {\tt
  if} statements at line~\ref{line:tan_cmov0}, \ref{line:tan_cmov1},
\ref{line:tan_cmov2}, and \ref{line:tan_cmov3} with
conditional move operators. The {\tt for} loop at line~\ref{line:tan_for}
is unrolled. We implemented {\tt round} functions at line~\ref{line:tan_round0},
\ref{line:tan_round1} and \ref{line:tan_round2}
with a single instruction for most of the vector
extensions. Macro ESTRIN at line~\ref{line:tan_estrin} evaluates a
polynomial in double-precision with Estrin's scheme.

In the loop from line~\ref{line:ph_forloopbegin} to
\ref{line:ph_forloopend} in the Payne-Hanek reduction, Eq.~(\ref{eq:ph0})
is computed. The result is multiplied by $\pi/2$ at
line~\ref{line:ph_mulpio2}. The numbers of FP operators shown in Table~\ref{table:paynehanek}
are the numbers of operators from line~\ref{line:ph_begin} to
line~\ref{line:ph_end}. The path with
the Payne-Hanek reduction is also taken if the argument is non-finite. In
this case, variable $x$ is set to NaN at line~\ref{line:tan_cmov1},
and this will propagate to the final result.

\section{Annotated source code of the FP remainder}

\label{sec:sourcecodexfmod}

\begin{figure}[tb]
\begin{tabular}{c}
\begin{lstlisting}
double xfmod(double x, double y) {
  double n = fabs(x), d = fabs(y), s = 1, q; /*\label{line:xfmod_abs}*/
  if (d < DBL_MIN) { /*\label{line:xfmod_if0}*/
    n *= 1ULL << 54;
    d *= 1ULL << 54;
    s = 1.0 / (1ULL << 54);
  }
  double2 r = dd(n);
  double rd = nextafter(1.0 / d, 0); /*\label{line:xfmod_rde}*/
  
  for(int i=0;i < 21 && r.x >= d;i++) { /*\label{line:xfmod_for}*/
    q = trunc(nextafter(r.x, 0) * rd); /*\label{line:xfmod_q}*/
    q = (3*d > r.x && r.x > d) ? 2 : q; /*\label{line:xfmod_case3}*/
    q = (2*d > r.x && r.x > d) ? 1 : q; /*\label{line:xfmod_case2}*/
    q = r.x == d ? (r.y >= 0 ? 1 : 0) : q;
    r = ddadd2_d2_d2_d2(r, ddmul_d2_d_d(q, -d)); /*\label{line:xfmod_r}*/
    r = ddnormalize_d2_d2(r);
  }
  
  double ret = r.x * s;
  if (r.x + r.y == d) ret = 0;  /*\label{line:xfmod_if1}*/
  ret = copysign(ret, x); /*\label{line:xfmod_copysign}*/
  if (n < d) ret = x;  /*\label{line:xfmod_if2}*/
  return d == 0 ? NAN : ret;
}
\end{lstlisting}
\end{tabular}
\caption{C source code of the FP remainder function}
\label{fig:xfmod}
\end{figure}

\begin{table}[tb]
\begin{center}
\caption{DD Functions}
\label{table:ddfunc}
\begin{tabular}{l|l}
\hline
Function name & Output\\
\hline
dd$(x, y)$ & DD number $x + y$ \\
ddadd2\_d2\_d2\_d2$(x, y)$ & Sum of DD numbers $x$ and $y$ \\
ddadd\_d2\_d2\_d2 & Addition of two DD numbers $x$ and $y$, \\
& where $|x| \geq |y|$\\
ddadd\_d2\_d\_d & Addition of two DP numbers $x$ and $y$, \\
& where $|x| \geq |y|$\\
ddadd\_d2\_d\_d2 & Addition of DP number $x$ and\\
&  DD number $y$, where $|x| \geq |y|$\\
ddmul\_d2\_d2\_d2$(x, y)$ & Product of DD numbers $x$ and $y$ \\
ddmul\_d2\_d2\_d$(x, y)$ & Product of DD number $x$ and\\
& DP number $y$\\
ddmul\_d2\_d\_d$(x, y)$ & Product of DP numbers $x$ and $y$\\
dddiv\_d2\_d2\_d2$(x, y)$ & Returns $x / y$, where $x$ and $y$ are DD\\
& numbers\\
ddsqu\_d2\_d2$(x)$ & Returns $x^2$, where $x$ is a DD number\\
ddscale\_d2\_d2\_d$(x, y)$ & Product of DD number $x$ and DP\\
& number $y$, where $y = 2^N$\\
ddnormalize\_d2\_d2$(x)$ & Re-normalize DD number $x$\\
\hline
\end{tabular}
\end{center}
\end{table}

Fig.~\ref{fig:xfmod} shows a sample C source code of our FP remainder. In
this implementation, both dividend and divisor are DP numbers. It correctly handles signs and
denormal numbers. It executes a division instruction
only once throughout the computation of the remainder. The
implementation also supports the case where a division operator does
not give a correctly rounded result. In this case, {\changed the} {\it nextafter}
function must be applied multiple times at line~\ref{line:xfmod_q}
according to the maximum error. We implemented {\tt if} statements at
line~\ref{line:xfmod_if0}, \ref{line:xfmod_if1} and
\ref{line:xfmod_if2} with conditional move operators.
In the main loop, the algorithm finds
the remainder of $n / d$, and at line~\ref{line:xfmod_copysign}, the
correct sign is assigned to the resulting remainder.

The stopping condition $r.x \geq d$ of the {\it for} loop at
line~\ref{line:xfmod_for} is not strictly required, and this loop can
be terminated after all the values in vectors satisfy the stopping
condition in a vectorized implementation.  Since we assume that all
operators return a round-to-nearest FP number, we use the {\it
  nextafter} function at line~\ref{line:xfmod_rde} and
\ref{line:xfmod_q} to find a value close to $r/d$ but not exceeding
it. This method is applicable only if $x/y$ is smaller than {\tt
  DBL\_MAX}. In order to find the correct $q$ when $r/d$ is between 1
and 3, we use a few comparisons to detect this case at line~\ref{line:xfmod_case3}
and \ref{line:xfmod_case2}.

\section{Correctness of FP Remainder}

\label{sec:prooffmod}

It is obvious that Algorithm~\ref{algorithm:fmod} returns a correct
result. We show partial proof that $r_k$ is representable as a
radix-2 FP number of precision $2p$ if $n$, $d$ and $q_k$ are radix-2
FP numbers of precision $p$.

We now define property $FP_p$ and function $RD_p$. $FP_p(x)$ holds iff
there are integer $0 \leq m < 2^p$ and integer $e$ that satisfy $x = m
\cdot 2^e$. If $x$ and $y$ are finite DP numbers, $FP_{53}(x)$ and
$FP_{106}(xy)$ holds. $RD_p(x)$ denotes the maximum number that does
not exceed $x$ and $FP_p(x)$ holds.

We now show that $FP_{2p}(r_k)$ and $FP_{2p}(q_kd)$ hold if $FP_p(n)$
and $FP_p(d)$ hold. By Lemma 1, integer $q_k$ exists that satisfies
$FP_p(q_k)$. Then, $FP_{2p}(r_0)$ and $FP_{2p}(q_kd)$
hold. $FP_{2p}(r_k-q_kd)$ holds because $r_k/2 \leq q_kd \leq
2r_k$. Then, $FP_{2p}(r_{k+1})$ holds.

\bigskip

\noindent
{\bf Lemma 1} Given $s \geq 1$ and integer $p \geq 2$. There exists an
integer $q$ that satisfies $FP_p(q)$ and $s/2 \leq q \leq s$.

\noindent
{\bf Proof:} If $s < 2^p$, $FP_p(\lfloor s \rfloor)$ holds. In this
case, $q$ can be $\lfloor s \rfloor$ since $\lfloor s \rfloor \geq
1$. Otherwise, $RD_p(s)$ is an integer. $(s - RD_p(s)) / s < 2^{1-p}$,
and therefore $(1 - 2^{1-p})s < RD_p(s)$. Because $p \geq 2$, $s/2 <
RD_p(s)$, and therefore $q$ can be $RD_p(s)$.

\bigskip

We now discuss the number of iterations. We suppose $q_k$ is set to
$\text{min}(RD_p(r_k/d), \lfloor r_k/d \rfloor)$. If $r_k/d < 2^p$,
$\lfloor r_k/d \rfloor=q_k$ and the loop terminates after $k$-th
iteration. Otherwise, $q_k = RD_p(r_k/d) < \lfloor r_k/d \rfloor$, and
$(r_k/d-q_k)/(r_k/d) < 2^{1-p}$.  Then, $r_{k+1} = r_k-q_kd < 2^{1-p}
r_k$. Therefore the number of iterations is $\lceil (\log_2(n/d))/(p-1)
\rceil$.

In order to show that the number that substitutes $q$ at
line~\ref{line:xfmod_q} in Fig.~\ref{fig:xfmod} satisfies the
condition at line~\ref{line:q_k} in Algorithm~\ref{algorithm:fmod}, we
prove $(r/d)/2 < \lfloor RN(nextafter(RN(r), 0)\cdot
nextafter(RN(1.0/d),0) \rfloor$ assuming $p=53$, $r/d \geq 3$ and $d >
0$. $RN(x)$ is the floating-point number that is the closest to $x$. 
We assume no overflow or underflow.

It is obvious that $q$ is substituted with a number that is smaller or
equal to $r/d$.  $(r - nextafter(RN(r), 0))/r < 2^{2-p}$, where $p$ is
precision. Therefore, $(1-2^{2-p})\cdot r < nextafter(RN(r),
0))$. Similarly, $(1-2^{2-p})\cdot 1.0/d < nextafter(RN(1.0/d), 0)$.
{Therefore,} $(1-2^{2-p})^2 (1-2^{1-p}) r/d < RN(nextafter(RN(r), 0)\cdot
nextafter(RN(1.0/d),0)$. Let $u = (1-2^{2-p})^2 (1-2^{1-p})$. $ur/d-1
\leq \lfloor ur/d \rfloor$. $(r/d)/2 < ur/d-1 \leftrightarrow 1 / (u -
1/2) < r/d$. $1 / (u - 1/2) < r/d$ holds since $p=53$ and $r/d \geq
3$. Therefore, $(r/d)/2 < \lfloor RN(nextafter(RN(r), 0)\cdot
nextafter(RN(1.0/d),0) \rfloor$.




\ifCLASSOPTIONcompsoc
  \section*{Acknowledgments}
\else
  \section*{Acknowledgment}
\fi

We wish to acknowledge Will Lovett and Srinath Vadlamani for their
valuable input and suggestions. The authors are particularly grateful
to Robert Werner, who reviewed the final manuscript.
The authors would like to thank Prof. Leigh McDowell for his suggestions.
The authors would like to thank all the developers that contributed to the SLEEF project, in particular Diana Bite and Alexandre Mutel.

\ifCLASSOPTIONcaptionsoff
  \newpage
\fi



%

\bibliography{tpds-sleef}
\bibliographystyle{IEEEtran}

\begin{IEEEbiography}{Naoki Shibata}
is an associate professor at Nara Institute of Science and
Technology. He received the Ph.D. degree in computer science from
Osaka University, Japan, in 2001. He was an assistant professor at
Nara Institute of Science and Technology 2001-2003 and an associate
professor at Shiga University 2004-2012. His research areas include
distributed and parallel systems, and intelligent transportation
systems. He is a member of IPSJ, ACM and IEEE.
\end{IEEEbiography}

\begin{IEEEbiography}{Francesco Petrogalli}
  is a software engineer working on the development of Arm Compiler
  for HPC. He contributed to the implementation of the Vector Length
  Agnostic (VLA) vectorizer for the Scalable Vector Extension (SVE) of
  Arm, to the ABI specifications for the vector functions on AArch64,
  and to the open source library SLEEF. Francesco has also worked on
  optimizing a variety of computational kernels that are core to
  Machine Learning algorithms.
\end{IEEEbiography}




\end{document}